\documentclass[aps,longbibliography,showpacs,twocolumn,superscriptaddress,amsmath,amssymb,verbatim]{revtex4-1}

\usepackage[thinlines]{easytable}
\usepackage{graphicx}
\usepackage{lmodern}
\usepackage{epstopdf}
\usepackage{dcolumn}
\usepackage{bm}
\usepackage{subfigure}
\usepackage{makecell}
\usepackage{amsmath} 
\usepackage[pdftex,colorlinks=true,citecolor=blue,linkcolor=blue,urlcolor=blue,bookmarks=true]{hyperref}
\usepackage{booktabs}
\usepackage{multirow} 
\usepackage{array} 

\usepackage{color}
\usepackage{textcomp}
\definecolor{red}{rgb}{1,0,0}

\definecolor{blue}{rgb}{0,0,1}

\definecolor{green}{rgb}{0,1,0}

\usepackage{multirow}
\usepackage[utf8]{inputenc}
\begin{document}
	\preprint{APS}

\title{Possible Bose-Einstein condensation of magnons in a $S$ = 5/2 honeycomb lattice  
 }

\author{J. Khatua}
\affiliation{Department of Physics, Sungkyunkwan University, Suwon 16419, Republic of Korea}
\author{S. M. Kumawat}
\affiliation{Department of Physics and Center for Quantum Frontiers of Research and Technology (QFort), National Cheng Kung University, Tainan, Taiwan}
\author{G. Senthil Murugan}
\homepage{nanosen@gmail.com}
\affiliation{Institute of Physics, Academia Sinica, Taipei 11529, Taiwan}	
\affiliation{Department of Physics, St. Joseph's College of Engineering, OMR, Chennai 600 119, India}
  \author{C.-L. Huang} 
\affiliation{Department of Physics and Center for Quantum Frontiers of Research and Technology (QFort), National Cheng Kung University, Tainan, Taiwan}
\author{Heung-Sik Kim}
\affiliation{
	Department of Semiconductor Physics and Institute of Quantum Convergence Technology, Kangwon National University, Chuncheon 24341, Republic of
	Korea}
\affiliation{Department of Energy Technology, Korea Institute of Energy Technology, Naju-si 58217, Republic of Korea}
\author{K. Sritharan}
\affiliation{Institute of Physics, Academia Sinica, Taipei 11529, Taiwan}
     \author{R. Sankar}
\homepage{sankarndf@gmail.com}
\affiliation{Institute of Physics, Academia Sinica, Taipei 11529, Taiwan}
\author{Kwang-Yong Choi}
 \homepage{choisky99@skku.edu}
\affiliation{Department of Physics, Sungkyunkwan University, Suwon 16419, Republic of Korea}

\date{\today}

\begin{abstract}

Quantum magnets offer a unique platform for exploring exotic quantum phases and quantum phase transitions through external magnetic fields. A prominent example is the field-induced Bose–Einstein condensation (BEC) of magnons near the saturation field. While this behavior has been observed in low-spin systems, its realization in high-spin, quasi-two-dimensional magnets—where multiple on-site excitations are possible—remains exceptionally rare. Here, we report thermodynamic and density functional theory results on single crystals of the honeycomb-lattice antiferromagnet K$_{4}$MnMo$_{4}$O$_{15}$ with $S = 5/2$. The system undergoes a field-induced transition to a fully polarized state at the critical field $\mu_{0}H_{\rm s}$ = 6.4 T. Our results reveal \textcolor{black}{possible} thermodynamic signatures of magnon BEC, $T_{\mathrm{N}} \sim (H_{\rm s} - H)^{2/d}$ ($d = 3$), expanding the purview of BEC-driven quantum criticality to a high-spin, quasi-two-dimensional antiferromagnets with negligibly small anisotropy.

\end{abstract}
\maketitle
\section{Introduction} 
Low-dimensional quantum magnets provide a fertile platform for realizing exotic quasiparticle excitations and exploring their many-body collective behavior~\cite{Balents2010,RevModPhys.86.563,PhysRev.126.1691,Wulferding2020,Kasprzak2006,Fogh2024,KHATUA20231}. External stimuli, particularly magnetic fields near the saturation point, can further induce novel quantum phases, resulting in rich phase diagrams with unconventional field-induced states including fractional magnetization plateaus~\cite{PhysRevLett.114.227202,PhysRevLett.82.3168}, Bose–Einstein condensation (BEC)~\cite{Giamarchi2008,RevModPhys.86.563,PhysRevLett.84.5868}, spin-nematic~\cite{andreev1984spin,PhysRevLett.118.247201,PhysRevLett.110.077206,pnas.1821969116}, and supersolid phases~\cite{Xiang2024,PhysRevLett.98.227201}.\\
 Among quasiparticle excitations in magnetic systems, bosonic spin excitations such as magnon and triplon can, under suitable conditions, condense into a single quantum state—giving rise to the phenomenon known as BEC under magnetic fields \cite{Giamarchi2008,RevModPhys.86.563}. This field-induced condensation represents a quantum phase transition and serves as a paradigmatic example of quantum criticality in some magnetic systems \cite{PhysRevLett.84.5868,Regg2003}. In earlier decades, aside from ultracold atomic gases,  three-dimensional (3D) dimerized antiferromagnets—such as TlCuCl\(_3\), BaCuSi\(_2\)O\(_6\), and Pb\(_2\)V\(_3\)O\(_9\)—have been extensively studied as platforms for triplon BEC \cite{PhysRevLett.96.077204,PhysRevLett.84.5868,PhysRevLett.93.087203,Greiner2002,PhysRevB.81.132401}. In these magnets an applied magnetic field closes the singlet–triplet gap, giving rise to field-induced \( XY \)-type magnetic order characterized by a low density of bosons \cite{Sebastian2006}. However, in such 3D magnets, tuning the boson density via the Zeeman energy to approach the BEC quantum critical point (BEC-QCP) remains relatively unexplored due to the experimental challenges of reaching high magnetic fields required to overcome strong intradimer interactions \cite{PhysRevB.79.100409,RevModPhys.86.563,Vasiliev2018}. \\
 \textcolor{black}{In addition to  BEC of triplons, conventional 3D and quasi-2D antiferromagnets exhibit field-induced saturation transitions, providing a versatile platform to study magnon BEC across a broad range of spin systems~\cite{RADU2007406,PhysRevLett.95.127202,PhysRevB.72.104414,DEJONGH1985737}. Within a widely discussed universal framework, the saturation field $H_{s}$
 	marks the onset of magnon BEC, with the transition temperature scaling as  $T_N \propto (H_s - H)^{\alpha}$ with $\alpha = 2/3$, in agreement with mean-field predictions and independent of the spin quantum number~\cite{PhysRevLett.95.127202,PTP.16.569,batyev1984antiferrornagnet}. While this universality is well established in many 3D spin systems, its validity in the quasi-2D limit is more subtle, with stability often strongly influenced by magnetic anisotropy.}\\ 
Noticeably, magnon condensation in 2D ordered magnets initially received little attention in the context of BEC. Moreover, BEC in 2D systems was long considered unattainable due to the finite density of states at zero energy; instead, a Berezinskii–Kosterlitz–Thouless transition is typically expected near the saturation field \cite{PhysRevLett.17.1133,berezinskii1972destruction,JMKosterlitz1972}.  However, recent observations of magnon BEC \textcolor{black}{with a exponent $\alpha = 1$,} in quasi-2D magnets with weak interplanar exchange interactions and easy-axis anisotropy have opened new avenues for exploring BEC-driven quantum criticality near the field-polarized phase, where 2D physics still dominates \cite{Matsumoto2024}. For example, the condensation of two-magnon bound state at the BEC-QCP driven by magnetic field and easy-axis anisotropy has been proposed  in a nearly perfect $S$ = 1 triangular lattice  Na$_{2}$BaNi(PO$_{4}$)$_{2}$ \cite{Sheng2025}. This system is particularly notable due to its weak exchange interactions, which allow the full temperature–magnetic field phase diagram to be mapped within a low field range of  2 T, revealing the elusive spin-nematic phase in proximity to the BEC-QCP \cite{Sheng2025}. Another example of magnon BEC in a quasi-2D honeycomb lattice is YbCl\(_3\) ($J_{\rm eff}$ = 1/2), which lies close to the 2D limit, with an interplanar-to-intraplanar coupling ratio of \( J_{\perp}/J = 2 \times 10^{-3} \) \cite{Matsumoto2024}.\\
In large-spin systems ($S>1$), the presence of multiple internal spin levels on each site allows for a variety of excitation pathways, enabling richer collective behavior and potentially more complex forms of BEC \cite{Prfer2022}. A rare realization of a field-induced double BEC dome has been reported  in the square-lattice compound Ba$_2$Co$_{1-x}$Zn$_x$Ge$_2$O$_7$ ($S$ = 3/2; $x = 0.25$), where the first dome arises from condensation within the lowest spin doublet ($|\pm\frac{1}{2}\rangle$), while the second emerges at higher magnetic fields, where a level crossing between the $\left|+\tfrac{1}{2}\right\rangle$ and $\left|+\tfrac{3}{2}\right\rangle$ states enables a secondary magnon condensation \cite{Watanabe2023}.\\\\
Motivated by the recent renewed interest in magnon BEC in 2D systems and the intriguing question of \textcolor{black}{how dimensionality and magnetic anisotropy can critically shape the character of the BEC transition across a wide variety of spin systems}, we explore the possibility of realizing BEC in  single crystals of a honeycomb-lattice compound K$_{4}$MnMo$_{4}$O$_{15}$ (hereafter KMMO), where Mn$^{2+}$ ions with spin $S = 5/2$ form a 2D honeycomb network perpendicular to the crystallographic $c$-axis. In zero field, KMMO undergoes long-range magnetic ordering at $T_{\rm N} = 2.21$ K, evidenced by a $\lambda$-like anomaly in the specific heat and in the temperature derivative of magnetic susceptibilities. Density functional theory (DFT) calculations reveal small intraplanar antiferromagnetic interactions of $J_{1}$ = 1.038 K, along with much weaker interplanar and second-neighbor intraplanar couplings, consistent with a negative Curie–Weiss temperature. Upon applying a magnetic field perpendicular to the $ab$-plane, the system evolves from Heisenberg-type order toward $XY$-like behavior, eventually entering a field-polarized phase beyond the critical field of $\mu_0 H_{\rm s} = 6.4$~T. At this transition, critical scaling of thermodynamic quantities provides signatures of the realization of a BEC-QCP in the 3D limit.

 \begin{figure*}
	\centering
	\includegraphics[width=\textwidth]{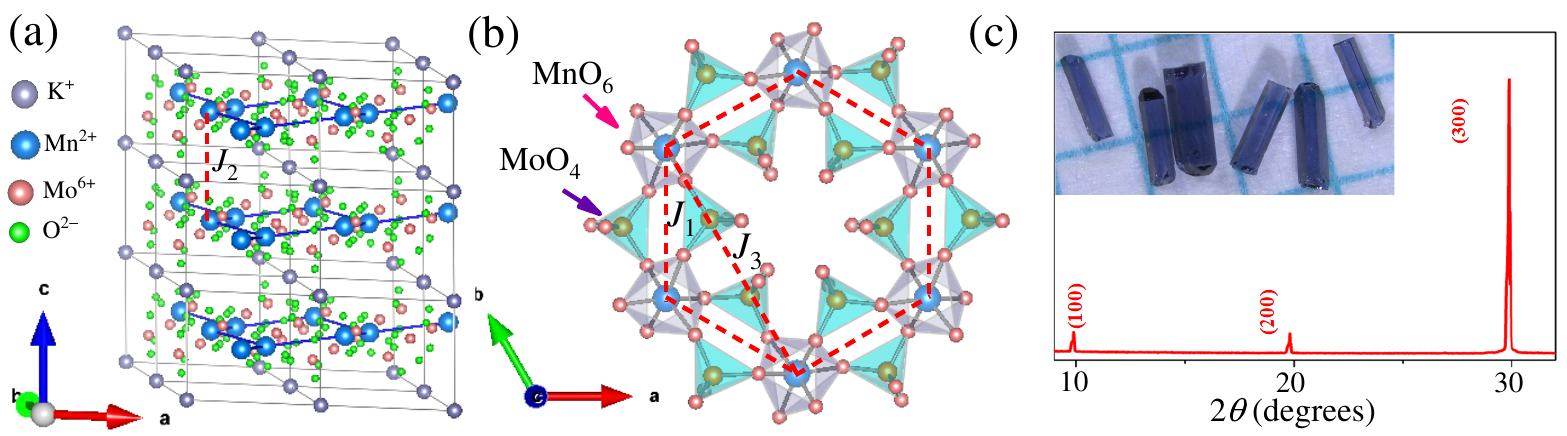}
	\caption{(a) Schematic of unit cells of K$_{4}$MnMo$_{4}$O$_{15}$, where Mn$^{2+}$ ions form a honeycomb lattice stacked along the $c$-axis. The interplanar interaction is indicated by the red dotted line labeled as $J{2}$. (b) Perpendicular view of the honeycomb plane formed by octahedrally coordinated Mn$^{2+}$ ions, where each Mn$^{2+}$ ion is connected to its nearest neighbor via the MoO$_4$ tetrahedra. The intraplanar  nearest- and next-nearest-neighbor interactions are labeled as $J_1$ and $J_3$, respectively.  (c) Powder X-ray diffraction pattern of a single crystal which has a preferred orientation of indexed (\textit{h}00) peaks. Inset shows the optical images of single crystals of K$_4$MnMo$_4$O$_{15}$. }{\label{KMMOFig1}}.
\end{figure*} 
\section{Experimental and theoretical details}
Polycrystalline samples of KMMO have been prepared by the standard solid-state reaction method. High-purity starting materials of K$_2$CO$_3$, MnO, MoO$_3$ (with a purity of 99.95$\%$) were thoroughly mixed and ground. The mixture was calcined in air at 400$^\circ$C for 48 hours in a ceramic crucible. To achieve a single-phase compound, the sample was further sintered at 450$^\circ$C and 500$^\circ$C for 48 hours, with intermediate grindings.\\
For single crystal growth, the polycrystalline powder was heated to 550$^\circ$C at a rate of 100$^\circ$C/h, held at 550$^\circ$C for 48 hours, then cooled to 525$^\circ$C and 500$^\circ$C at rate of 0.5$^\circ$C/h, and finally cooled
down to room temperature with a cooling rate of 50$^\circ$C/h. Deep blue crystals ($\sim 5 \times 2 \times 2$ mm³) were successfully grown and mechanically collected from the crucible.
\\
 Powder X-ray diffraction (XRD) measurements were carried out at room temperature on crushed single crystals of KMMO using a Bruker AXS D8 Advance diffractometer with Cu K$\alpha$ radiation ($\lambda = 1.54$~\AA). The diffraction pattern was obtained with with the X-ray beam aligned perpendicular to the (\textit{h}00) planes of a single crystal using the same setup. \\ Magnetic measurements were carried out  using a superconducting quantum interference device vibrating-sample magnetometer (SQUID-VSM, Quantum Design, USA) in the temperature range 2 K $\leq$ $T$ $\leq$ 300 K and in magnetic fields up to 7 T. Additional magnetization measurements were performed  using the $^{3}$He option of the MPMS3 SQUID magnetometer from Quantum Design. Specific heat measurements were conducted using a standard relaxation method with a physical property measurement system (PPMS, Quantum Design, USA) in the temperature range 0.13 K $\leq$ $T$ $\leq$ 300 K in several magnetic fields up to 9 T.   \\ DFT calculations were carried out using the OpenMX code with the Perdew–Burke–Ernzerhof (PBE) generalized gradient approximation (GGA) as the exchange-correlation functional \cite{PhysRevB.67.155108}. To account for strong electronic correlations in Mn atoms, a Hubbard on-site Coulomb parameter $U$ was applied with varying values. The plane-wave energy cutoff was set to 300 Ry, and self-consistent field (SCF) convergence was achieved with a threshold of $1.0 \times 10^{-9}$ Hartree. The Brillouin zone was sampled using a $4 \times 4 \times 6$ Monkhorst–Pack $k$-point mesh. SCF iterations employed the RMM-DIIS mixing scheme, with a maximum of 300 steps and adaptive mixing parameters.\\
The converged SCF results from OpenMX served as input for the \textsc{J$_{X}$} code \cite{YOON2020106927}, which evaluates exchange coupling parameters \( J^{\text{GGA}}_{ij} \) between localized spins based on the Green's function formulation of the Liechtenstein approach.\\
\section{Results and discussion}
\subsection{Crystal structure} To confirm the crystal structure of KMMO, Rietveld refinement of powder XRD data--obtained from crushed single crystals--was carried out at room temperature. The refinement results indicate that the title compound KMMO crystallizes in a trigonal structure (space group \textit{P}\,-3) with the lattice parameters $a = b = 10.37$~\AA, $c = 8.16$~\AA, and angles $\alpha = \beta = 90^\circ$, $\gamma = 120^\circ$.  The obtained lattice parameters and atomic coordinates (not shown here) are consistent with earlier reported values  \cite{Solodovnikov1997}. Figure~\ref{KMMOFig1}(a) shows a schematic of several unit cells of KMMO, where the magnetic Mn$^{2+}$ ions occupy a unique crystallographic site, without any detectable anti-site disorder among the constituent ions. Interestingly, the Mn$^{2+}$ ions form a nearly perfect 2D honeycomb lattice with a nearest-neighbor distance of 6.01~\AA, oriented perpendicular to the $c$-axis. Notably, the interplanar Mn–Mn distance (8.16~\AA) is shorter than the intraplanar second-nearest-neighbor distance (10.37~\AA). Each Mn$^{2+}$ ion is coordinated by six O$^{2-}$ ions, forming an MnO$_6$ octahedron. Two of these oxygen atoms are shared with the two adjacent MoO$_4$ tetrahedra, each of which shares one of its corners with the nearest-neighbor oxygen atoms of the MnO$_6$ octahedron (see Fig.~\ref{KMMOFig1}(b)). This arrangement establishes an intraplanar nearest-neighbor superexchange pathway mediated through the Mn–O–Mo–O–Mn connection. The XRD pattern obtained with the incident beam oriented perpendicular to the ($h$00) plane is shown in Fig.~\ref{KMMOFig1}(c), which corresponds to the orientation perpendicular to the honeycomb plane. The inset of Fig.~\ref{KMMOFig1}(c) displays photos of KMMO crystals with their top surfaces corresponding to the ($h$00) plane. \\ \\ 
\subsection{Thermodynamic properties and DFT calculations}\label{DFT}
\begin{figure}
	\centering
	\hspace{0cm}
	\includegraphics[width=0.5\textwidth]{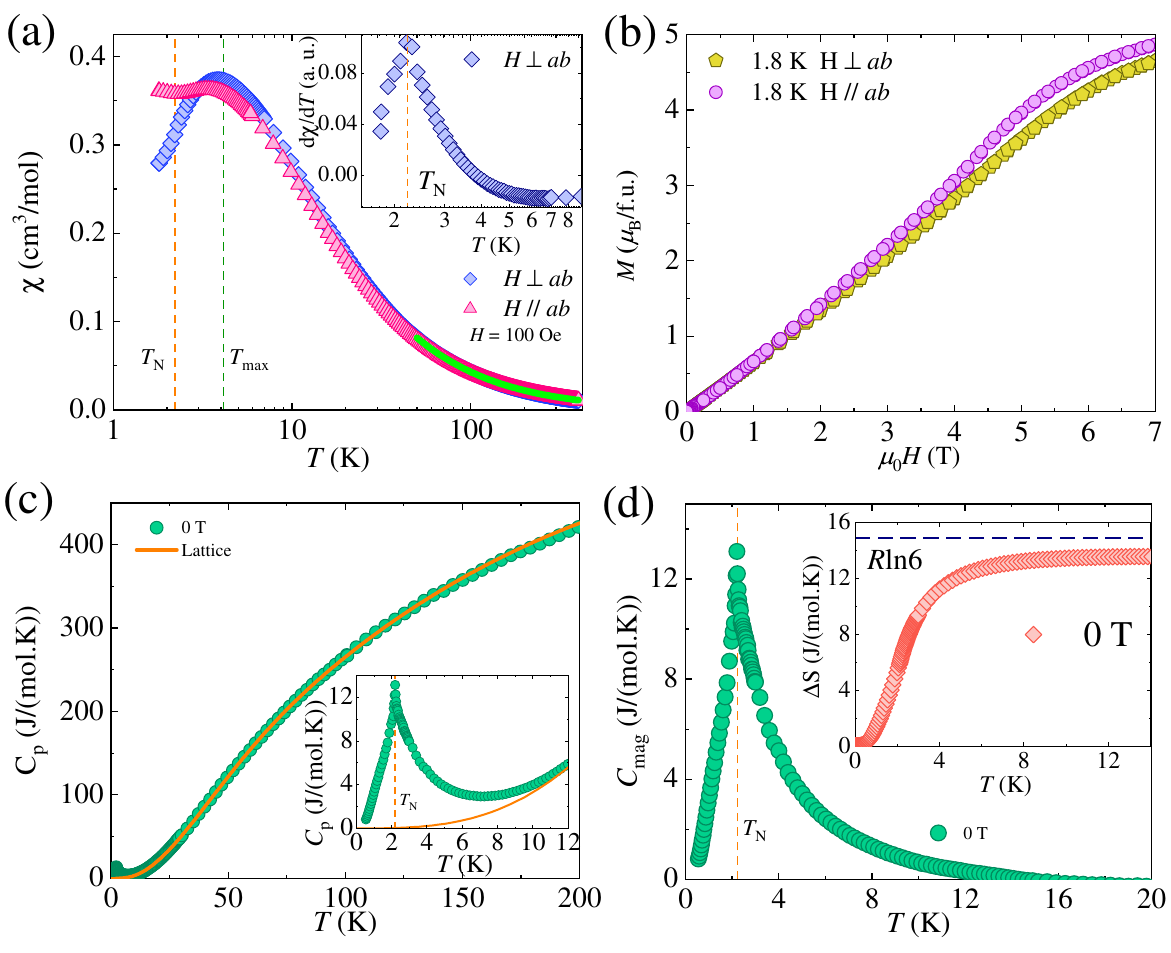}
	\caption{(a) Temperature dependence of magnetic susceptibility measured in a field of $\mu_0H = 0.01$~T applied parallel and perpendicular to the \textit{ab}-plane, with the $x$-axis plotted on a logarithmic scale. The solid green line represents the Curie–Weiss fit, while the dashed vertical lines indicate the broad maximum at $T_{\rm max} = 4.2$~K and the Néel temperature at $T_{\rm N} = 2.21$~K. The top inset displays the derivative of magnetic susceptibility as a function of temperature for the field applied perpendicular to the \textit{ab}-plane. (b) Isothermal magnetization as a function of magnetic field parallel and perpendicular to the \textit{ab}-plane at 1.8 K. (c) Temperature dependence of specific heat in  zero field, where the solid orange line represents the Debye–Einstein model of the lattice contribution. The bottom inset zooms into the low-temperature region, revealing an anomaly at $T_{\rm N}$. (d) Magnetic specific heat as a function of temperature showing an anomaly at $T_{\rm N}$. The top inset shows the temperature dependence of the calculated entropy change in zero field.      }{\label{KMMOFig2}}.
\end{figure}

In order to investigate the behavior of  local moments of Mn$^{2+}$ ($S = 5/2$) ions, their exchange interactions, and anisotropic properties, magnetic susceptibility ($\chi(T)$) measurements were performed in a field of $\mu_0H = 0.01$~T applied parallel and perpendicular to the $ab$-plane as shown in Fig.~\ref{KMMOFig2}(a).  Upon lowering the sample temperature, $\chi(T)$ exhibits no significant directional dependence at high temperatures; however, deviations between the two directions begin to emerge below 30~K, indicating the onset of magnetic correlations with moderate anisotropy. Above $T> 30$ K, the inverse susceptibility remains linear,  indicating the Curie–Weiss (CW) regime, and is well described by the CW law, $\chi(T) = \chi_{0} + C/(T - \theta_{\mathrm{CW}})$. Here, $\chi_0$ represents the temperature-independent contributions from core diamagnetism and Van Vleck paramagnetism, $C$ is the Curie constant, and $\theta_{\mathrm{CW}}$ reflects the strength and nature of magnetic exchange interactions. The green solid line in Fig.~\ref{KMMOFig2}(a) represents the CW fit, yielding $\chi_0 = -3.27 \times 10^{-4}$~cm$^3$/mol, $C = 4.7 \pm 0.01$~cm$^3$K/mol, and $\theta_{\rm CW} = -9 \pm 0.25$~K.  The calculated effective magnetic moment $\mu_{\rm eff}$ = $\sqrt{8C}$ = 6.13 $\mu_{\rm B}$ is slightly larger than the spin-only value of $\mu_{\mathrm{eff}} = 5.91~\mu_{\mathrm{B}}$
 for high-spin Mn$^{2+} (S = 5/2)$ \cite{Khatua2021}. The obtained negative CW temperature indicates that the dominant magnetic interactions between the $S = 5/2$ moments are antiferromagnetic in nature. \\ 
Upon further cooling below 30 K, $\chi(T)$ increases monotonically for both directions ($\chi_{\perp}$ and $\chi_{\parallel}$), reaching a broad maximum around $T_{\rm max} = 4.2$ K, indicating the presence of short-range spin correlations (SRO), typical of low-dimensional magnetic systems \cite{Vasiliev2018}. Below $T_{\rm max}$, $\chi_{\perp}$ begins to decrease, with a distinct change in slope across $T_{\rm N}$ = 2.21 K, as evidenced by the anomaly in $d\chi(T)/dT$ shown in the inset of Fig.~\ref{KMMOFig2}(a). In contrast, $\chi_{\parallel}$ continues to increase, displaying a weak dip near $T_{\rm N}$. The observation of $\chi_{\parallel} > \chi_{\perp}$ suggests the presence of easy-plane anisotropy. Figure~\ref{KMMOFig2}(b) shows the isothermal magnetization as a function of magnetic field, which tends toward saturation above 7~T, consistent with the estimated CW temperature. It is worth noting that the isothermal magnetization exhibits no noticeable anisotropy up to 1.6~T, beyond which a moderate anisotropic behavior begins to emerge. This crossover may be associated with the transition from Heisenberg-like  to $XY$-like antiferromagnetic behavior, as also supported by the field-dependent specific heat measurements (see below). \\ To further confirm the presence of long-range magnetic order, specific heat ($C_{\rm p}$) measurements were performed in  zero field as shown in Fig.~\ref{KMMOFig2}(c). A clear $\lambda$-like anomaly is observed around $T_{\rm N}$ (see inset of Fig.~\ref{KMMOFig2}(c)) which corresponds to the  anomaly observed in the $d\chi/dT$ data, further supporting the presence of long-range magnetic order in KMMO. To subtract the phonon contribution to the specific heat, the $C_p$ data were fitted (solid orange line in Fig.~\ref{KMMOFig2}(c)) using a model comprising one Debye term and three Einstein terms i.e.,
\( C_{\rm latt}(T) = C_D \left[ 9R \left(\frac{T}{\theta_D} \right)^3 \int_0^{\theta_D/T} \frac{x^4 e^x}{(e^x - 1)^2} \, dx \right] + \sum_{i=1}^{3} C_{E_i} \left[ R \left( \frac{\theta_{E_i}}{T} \right)^2 \frac{e^{\theta_{E_i}/T}}{(e^{\theta_{E_i}/T} - 1)^2} \right] \), where $\theta_{D}$ = 120 $\pm$ 0.30 K is the Debye temperature, $\theta_{E_1} = 177 \pm 0.54$ K, $\theta_{E_2} = 302 \pm 0.76$ K, $\theta_{E_3} = 671 \pm 1.6$~K are the Einstein temperatures of the three optical phonon modes, and \textit{R} is the molar gas.   To reduce the number of fitting parameters, $C_D$ was fixed at 3 to represent the three acoustic phonon modes, while $C_{E_1} = 15$, $C_{E_2} = 20$, and $C_{E_3} = 25$ were assigned to account for the 69 optical modes, corresponding to the (3$n$–3) optical branches for $n = 24$ atoms in KMMO \cite{PhysRevB.110.184402}. After subtraction of the lattice contributions, the resulting magnetic specific heat is shown in Fig.~\ref{KMMOFig2}(d), revealing an anomaly around $T_{\rm N}$ and indicating that magnetic correlations begin to develop at temperatures higher than the CW temperature, consistent with the $\chi(T)$ data. Next, the change of magnetic entropy associated to the magnetic ordering was calculated by integrating the magnetic specific heat divided by temperature as shown in the inset of Fig.~\ref{KMMOFig2}(d). The total entropy released above the CW temperature is about 13.65 J/mol·K, corresponding to roughly 91\text{\%} of the expected value, $R\ln(2S + 1)$, for  a $S = 5/2$ system. The missing 10\text{\%} of  the entropy might be due to the overestimation of lattice contribution or because of short-range spin correlations that exist above $T_{\rm N}$. Interestingly, only about 40\text{\%} of the total entropy is released at $T_{\rm N}$, meaning that the rest is released at higher temperatures due to short-range magnetic interactions. This agrees well with the broad maximum seen in the temperature dependence of the $\chi(T)$ data (Fig.~\ref{KMMOFig2}(a)). \\
\begin{figure}[b]
	\centering
	\includegraphics[width=0.45\textwidth]{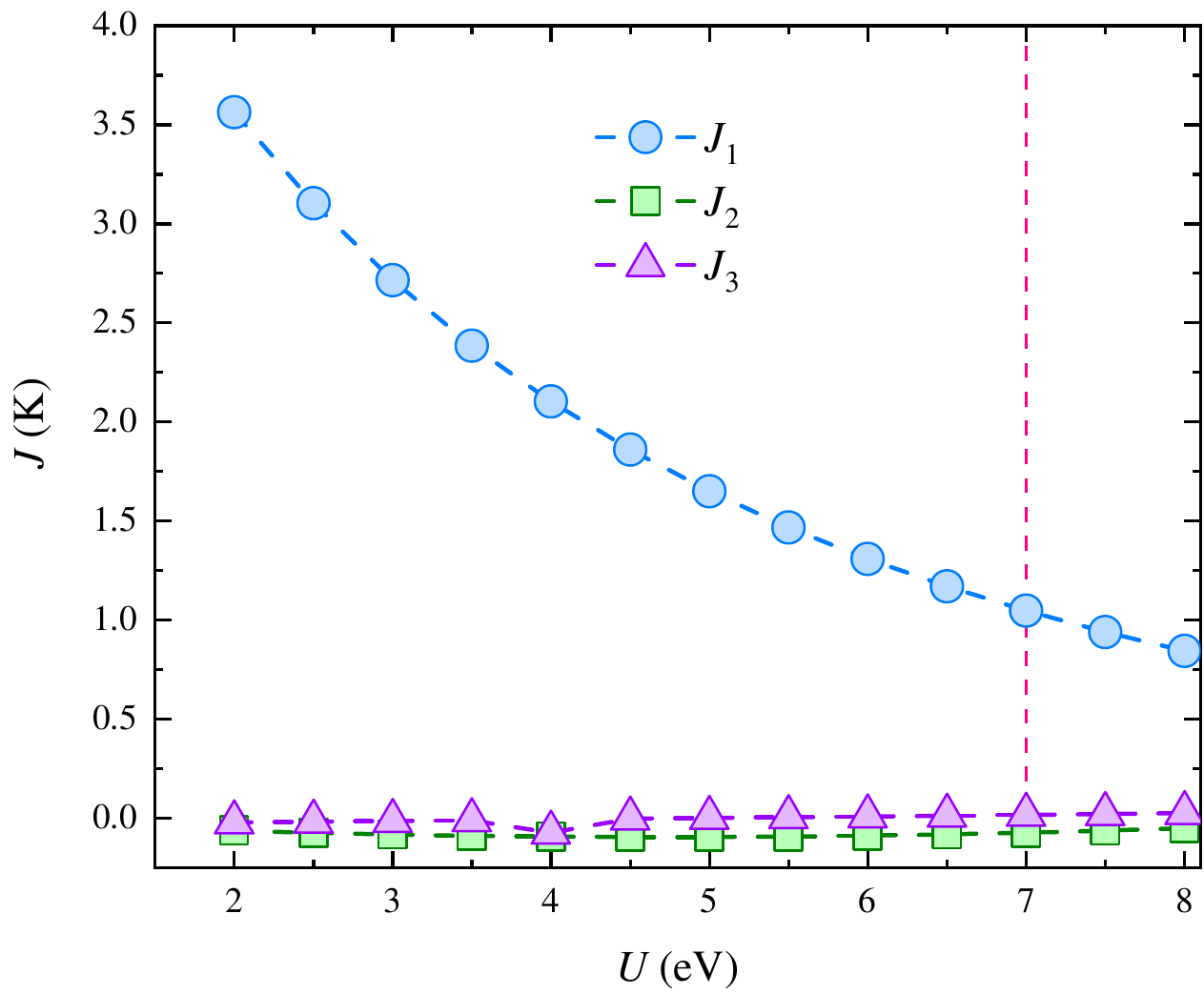}
	\caption{Variation of the three exchange interactions with respect to the on-site Coulomb interaction strength $U$. The vertical dashed pink line indicates the specific value for which the calculated exchange couplings reproduce the experimental Curie--Weiss temperature }{\label{KMMOFig3}}.
\end{figure}
 \begin{figure*}[t]
	\centering
	\hspace{0cm}
	\includegraphics[width=1\textwidth]{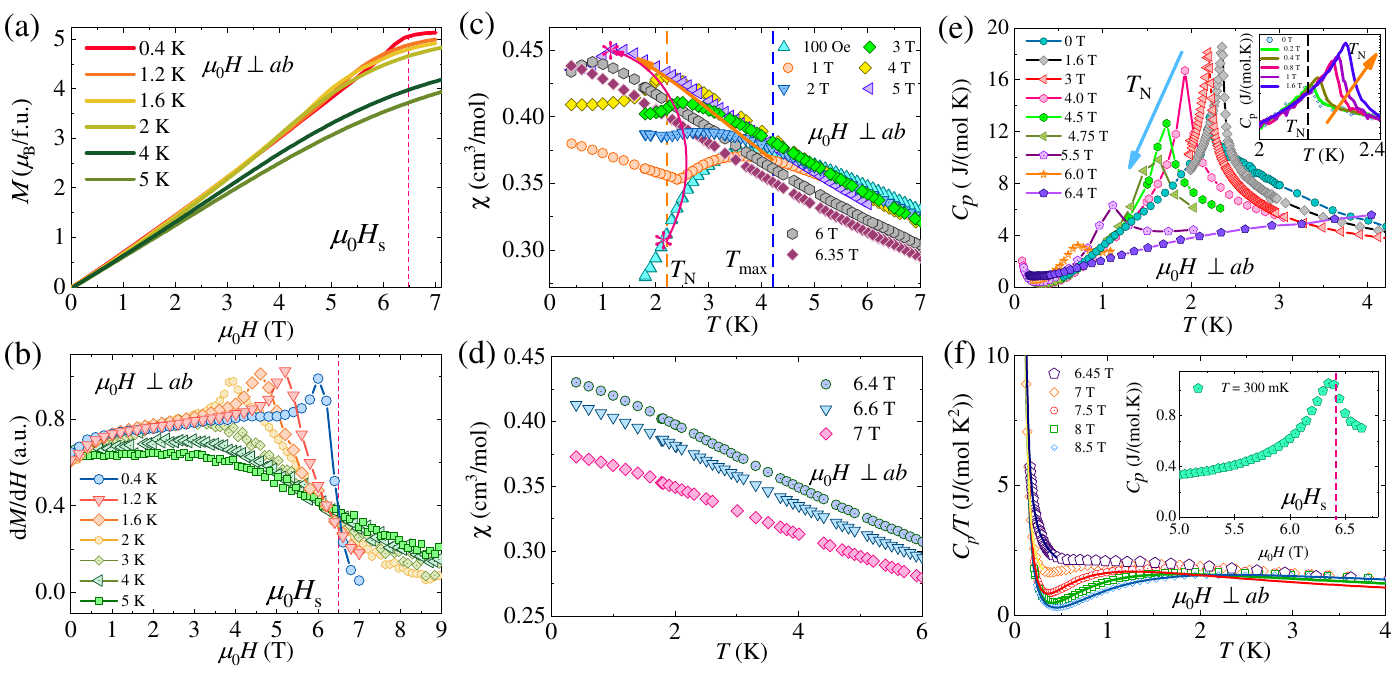}
	\caption{(a) Isothermal magnetization as a function of magnetic field and (b) its derivative. The dashed vertical lines indicate the position of  the quantum critical point at the saturation field $\mu_0H_{\rm s} = 6.4$\,T. (c) Temperature dependence of magnetic susceptibility at low temperatures under several magnetic fields. Dashed vertical lines indicate $T_{\rm N}$ and $T_{\rm max}$ at $\mu_0H = 0.01$\,T, with their field-dependent shifts highlighted by orange arrows (for $T_{\rm max}$) and pink star arrows (for $T_{\rm N}$). (d) Temperature dependence of magnetic susceptibility in fields $\mu_{0}H\geq 6.4$ T at low temperatures. (e) Temperature dependence of specific heat at low-temperatures in several fields. \textcolor{black}{The inset shows a magnified view of the anomaly up to 1.6 T.} The orange and sky-blue arrows indicate the progressive shifts of $T_{\rm N}$ toward higher and lower temperatures, respectively. (f) Temperature dependence of specific heat divided by temperature for fields $\mu_{0}{H}> 6.4$ T. The solid lines represent a combination of fits to the nuclear Schottky contribution and the gapped behavior, as described in the text.
		 Inset shows the field-dependent specific heat at $T$ = 300 mK. In all panels, the magnetic field was applied perpendicular to the \textit{ab}-plane.
	}{\label{KMMOFig4}}.
\end{figure*}
In order to determine a spin Hamiltonian of KMMO, we computed the interatomic exchange interactions using the magnetic force linear response theory~\cite{LIECHTENSTEIN198765}. This approach allowed us to quantify the strength of intraplanar nearest-neighbor ($J_1$) and second-nearest-neighbor ($J_3$) interactions, as well as the interplanar coupling ($J_2$) (see Fig.~\ref{KMMOFig1} (a) and (b)). Figure~\ref{KMMOFig3} presents the variation of these three exchange interactions as a function of the on-site Coulomb interaction $U$. It reveals the presence of  a dominant antiferromagnetic nearest-neighbor exchange interaction $J_1$ while $J_2$ and $J_3$ remain relatively weak. 
Using the calculated $J_1 = 1.038$~K, $J_{2}$ = $-0.073$ K, and $J_{3}$ = 0.0165 K for $U$ = 7 eV , the Curie–Weiss temperature was estimated using the relation  $\theta_{\rm CW} = S(S+1)(3J_1+2J_{2} + 6J_{3})/3$, yielding $|\theta_{\rm CW}| \approx 9.03$~K, in good agreement with the experimental value.
The weak second-neighbor exchange within the honeycomb plane underscores that the origin of \(T_{\rm N}\) being well below \(\theta_{\rm CW}\) is due to the low-dimensional nature of the system, while long-range magnetic order likely arises from a combination of interplanar coupling and weak easy-plane anisotropy.\\
\subsection{Field-induced thermodynamic properties}\label{FITP} To further explore the field-tunable magnetic ground state, thermodynamic measurements were carried out at several fields applied perpendicular to the \textit{ab}-plane. Figure~\ref{KMMOFig4}(a) presents the isothermal magnetization at several temperatures, while its field derivative is plotted in Fig.~\ref{KMMOFig4}(b). In the antiferromagnetic state below $T_{\rm N}$, magnetization responds linearly to the applied field up to 4~T. A steeper  rise at higher fields reflects a transition toward the fully polarized (FP) phase. The critical field ($\mu_{0}H_{\rm s}$) associated with this field-induced transition is identified by an anomaly in the field derivative of the magnetization (see Fig.~\ref{KMMOFig4}(b)). The dashed vertical line at $\mu_{0}H_{\rm s} = 6.4$~T indicates the quantum critical point at the saturation field for the titled compound KMMO, as determined from field-dependent specific heat measurements at $T = 0.3$~K (see below). As the temperature approaches $T_{\rm N}$, the field-induced anomaly in $dM/dH$ (Fig.~\ref{KMMOFig4}(b)) gradually disappears due to enhanced  thermal fluctuations.\\  \begin{figure*}
	\includegraphics[width=\textwidth]{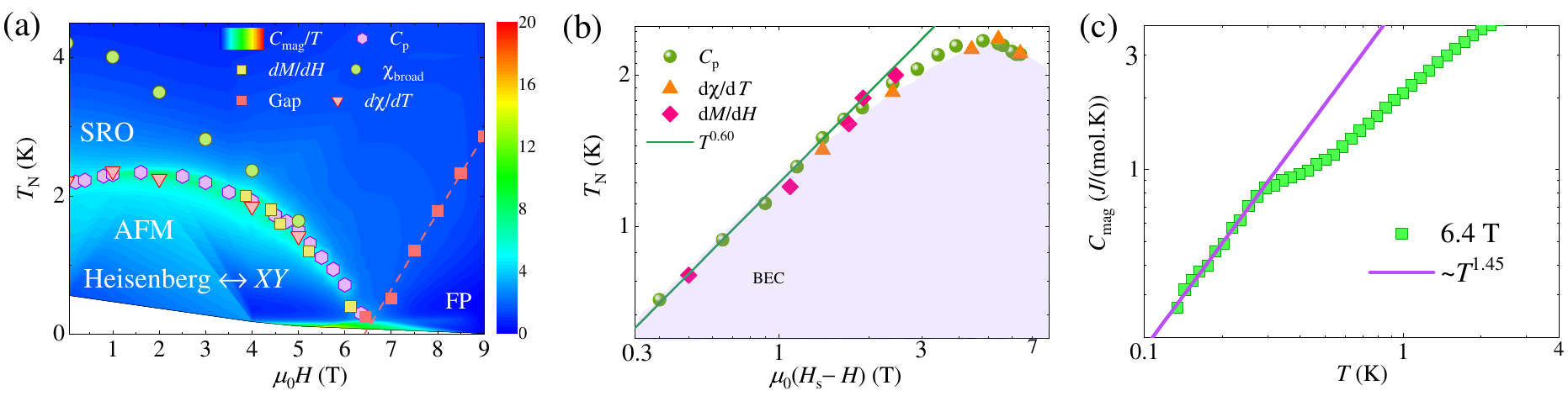}
	\caption{(a) Temperature–magnetic field phase diagram with phase boundaries determined from the thermodynamic measurements, as indicated in the legend. The background shows a contour map of the magnetic specific heat divided by temperature. The dashed red line shows a linear field dependence of gap.
		 (b) Scaling behaviour of $T_{\rm N}$ as a function of $\mu_{0}(H_{\rm s}-H)$ on a logarithmic scale. (c) Temperature dependence of magnetic specific heat at the critical field $\mu_{0}H_{\rm s}$ = 6.4 T on a logarithmic scale. The solid line indicates a $\sim T^{1.45}$ power-law behaviour.   }{\label{KMMOFig5}}.
\end{figure*}
Figure~\ref{KMMOFig4}(c) shows the temperature dependence of \(\chi(T)\) under several magnetic fields at low temperatures. The dashed vertical lines correspond to \(T_{\rm max} = 4.2\)~K, associated with SRO, and the Néel temperature \(T_{\rm N}\), determined from the anomaly in $d$\(\chi\)/$d$\(T\) (see Fig.~\ref{KMMOFig2}b) at \(\mu_{0}H = 0.01\)~T. As the magnetic field increases, \(T_{\rm max}\) shifts to lower temperatures (orange arrow), while \(T_{\rm N}\) initially shifts to higher temperatures (star pink arrow) before decreasing at higher fields. This non-monotonic behavior of \(T_{\rm N}\) suggests a crossover from Heisenberg to $XY$-type spin order under an applied magnetic field, which arises from the interplay between spin dimensionality, anisotropy, and Zeeman energy in near-isotropic Heisenberg magnets with small single-ion anisotropy. More specific, when a magnetic field is applied perpendicular to the easy plane, the magnetic field quenches the out-of-plane spin fluctuations, effectively leading to a dimensional reduction in spin space \cite{Matsumoto2024}.  For fields \(\mu_{0}H_{\rm s} \geq 6.4\)~T,  \(\chi(T)\) (Fig.~\ref{KMMOFig4}(d)) exhibits a monotonic increase with decreasing temperature, along with progressively reduced \(\chi(T)\) values with increasing field, consistent with a FP phase. In this regime, the external field aligns the magnetic moments, suppressing thermal spin fluctuations and reducing the system’s susceptibility to respond to further changes in field, resulting in lower $\chi(T)$.\\
To support the symmetry crossover observed in the \(\chi(T)\) data and to accurately determine the critical field, specific heat serves as a powerful probe. Figure~\ref{KMMOFig4}(e) presents the temperature dependence of \(C_p\) under several magnetic fields. With increasing field, the \(\lambda\)-like anomaly becomes more pronounced and shifts to higher temperatures up to 1.6~T (\textcolor{black}{as indicated by  the orange arrow in the inset of Fig.~\ref{KMMOFig4}(e)).} Beyond this field, the anomaly gradually moves to lower temperatures, consistent with a crossover from Heisenberg to $XY$-type spin order, as also reflected in the \(\chi(T)\) data. Interestingly, the \(\lambda\)-like anomaly is completely suppressed at the critical field $\mu_{0}H$$_{s}$ = 6.4 T. Below 0.5~K, the upturn in the specific heat is attributed to nuclear Schottky contributions, exhibiting the characteristic \(1/T^{2}\) dependence \cite{Matsumoto2024}. The presence of the critical field is further supported by an anomaly observed in the field-dependent \(C_p\) data at 0.3~K, as shown in the inset of Fig.~\ref{KMMOFig4}(f).  \\
To investigate the FP phase, \(C_p\) measurements were carried out at several fields above \(\mu_{0}H_{\rm s}\). Figure~\ref{KMMOFig4}(f) presents \(C_p/T\) as a function of temperature, revealing a broad maximum at higher temperatures and a pronounced upturn  at low temperatures, indicating the presence of a field-induced gap and nuclear Schottky contribution, respectively. The solid line in Fig.~\ref{KMMOFig4}(f) represents a fit using the model \(C_p(T) \propto 1/T^{2} + \exp(-\Delta/T)\) \cite{Matsumoto2024}, where \(\Delta\) represents the value of field-induced gap. The obtaned gap values are plotted in orange squares of Fig.~\ref{KMMOFig5}(a).\\
\subsection{Magnetic phase diagram and critical scaling behaviour}
To provide a comprehensive picture of the evolution of the field-induced phase transition from the antiferromagnetic (AFM) to the FP phase, we constructed the temperature–magnetic field phase diagram (Fig.~\ref{KMMOFig5}(a)) based on the thermodynamic results.  The different regions are labeled according to the interpretations discussed in Secs.~\ref{DFT} and \ref{FITP}. Above  \(\mu_0 H_s\), a gapped FP phase emerges. With increasing magnetic field, the estimated gap value follows a linear dependence of the form \(g\mu_{\rm B}(H - H_s)\) (Fig.~\ref{KMMOFig5}(a)), yielding \(g = 1.66\). \textcolor{black}{This effective $g$-value is somewhat lower than the typical $g$-factor $g$$\sim$2.0 for Mn$^{2+}$ ions.} The phase boundaries are plotted over the contour map of the magnetic specific heat divided by temperature.\\ To assess whether the field-induced phase transition near $\mu_{0}H_{\rm s}$ can be characterized as a BEC-QCP of magnons, we plotted \(T_{\rm N}\) as a function of $\mu_{0}$(\(H_s - H\)) \cite{RevModPhys.86.563}. Remarkably, the data follow a power-law behavior \(T_{\rm N} \propto (H_s - H)^{0.60 \pm 0.01}\) (Fig.~\ref{KMMOFig5}(b)), consistent with the BEC universality class in 3D $XY$-type antiferromagnets, where \(T_{\rm N} \sim (H_{\rm s} - H)^{2/d}\) ($d$ = 3) is expected \cite{Matsumoto2024,RevModPhys.86.563}. Additionally, the magnetic specific heat after subtracting the nuclear Schottky contribution exhibits a power-law dependence \(C_{\rm mag} \propto T^{1.45 \pm 0.02}\), aligning with the \(T^{d/2}\) behavior predicted for BEC in 3D dimensional systems (Fig.~\ref{KMMOFig5}(c)). These critical exponents of thermodynamic quantities strongly support the realization of the field-induced transition at $\mu_{0}H_{\rm s}$ as a BEC-QCP of magnons. \textcolor{black}{Although our initial data suggest the possibility of magnon BEC in the 3D limit, further low-temperature experiments in the vicinity of the critical field are required to confirm this behavior more unambiguously.} In addition, at $\mu_{0}H_{\rm s}$ , one might expect \( M_{\rm s} - M \propto T^{3/2} \); however, due to the limited field range, we are unable to reliably determine the saturation magnetization ($M_{\rm s}$). \\
In contrast to the BEC  observed in 2D systems such as the triangular lattice Na$_{2}$BaNi(PO$_{4}$)$_{2}$ (\(S = 1\)) \cite{Sheng2025} and the honeycomb lattice YbCl$_{3}$ (\(J_{\rm eff} = 1/2\)) \cite{Matsumoto2024} at the saturation field, the present quasi-two-dimensional $S$ = 5/2 system exhibits magnon BEC behavior  in the 3D limit, highlighting influence of both interlayer interactions and spin magnitude in governing the nature of quantum phase transitions. Comparatively, the $S=5/2$ honeycomb-lattice antiferromagnet FeP$_{3}$SiO$_{11}$ does not exhibit the BEC scenario, displaying markedly different characteristics from the present compound. \textcolor{black}{This distinction primarily arises from the stronger magnetic anisotropy intrinsic to Fe$^{3+}$-based systems compared with Mn$^{2+}$ magnets~\cite{mi9060292}, as reflected in the distinct $g$-values of $g_{1} \approx 2.018$ and $g_{2} \approx 2.001$ in FeP$_{3}$SiO$_{11}$~\cite{PhysRevB.110.184402}. In addition, although both systems feature comparable nearest-neighbor intraplanar couplings, FeP$_{3}$SiO$_{11}$ hosts two additional inequivalent antiferromagnetic interplanar interactions amounting to $\sim$ 16.7\% and 27.2\% of $J = 0.863$ K~\cite{PhysRevB.110.184402}, whereas KMMO exhibits weak ferromagnetic interplanar coupling ($J_{2}=-0.073$ K) alongside with dominant nearest-neighbor antiferromagnetic exchange interactions.  Notably, the sizable interplanar interactions in FeP$_{3}$SiO$_{11}$, relative to its intraplanar couplings, give rise to disparate magnetic correlations along the in-plane and out-of-plane directions. These contrasts are further manifested in their transition temperatures and the field-induced evolution of $T_{N}$, leading to distinct field–temperature phase diagrams in the two systems. 
	Our results thus demonstrate that the combined effects of magnetic anisotropy and interplanar interactions play a decisive role in determining whether a field-induced transition can stabilize BEC criticality, even in systems with the same spin quantum number.	}
	Moreover, the occurrence of BEC in magnets with higher spin number underscores the importance of single-site quantum level structure, where multiple spin projection states can enhance magnon interactions. Future high-frequency electron spin resonance experiments are called for to probe magnon bound states near the saturation field.
\vspace*{0.1 cm}

\section{CONCLUSION}
In summary, we have successfully synthesized single crystals and investigated the thermodynamic properties of a quasi-two-dimensional compound K$_{4}$MnMo$_{4}$O$_{15}$, where Mn$^{2+}$ ions with \( S = 5/2 \) form a honeycomb lattice perpendicular to the \( c \)-axis, supported by DFT calculations.  Our results establish KMMO as a rare example of a quasi-two-dimensional honeycomb-lattice antiferromagnet with \( S = 5/2 \), exhibiting field-tunable quantum critical behavior associated with magnon BEC. A $\lambda$-like anomaly in the derivative of magnetic susceptibilities and specific heat confirms the presence of long-range ordered state below $T_{\rm N}$ = 2.21 K, while DFT calculations suggest a dominant intraplanar antiferromagnetic exchange network with weaker interplanar and next-nearest-neighbor intraplanar couplings—consistent with the obtained Curie-Weiss temperature. Upon increasing the magnetic field perpendicular to the \( ab \)-plane, the system exhibits a crossover from Heisenberg to $XY$ anisotropy, followed by a transition to a fully polarized state across the critical field \( \mu_0 H_{\rm s} \) = 6.4 T. The observed critical exponent near \(\mu_{0} H_{\rm s} \) highlights the \textcolor{black}{possible} realization of a Bose–Einstein condensation quantum critical point  in this higher-spin honeycomb lattice. Our study therefore broadens the landscape of BEC associated quantum criticality into a quasi-2D honeycomb lattice with higher spin degrees of freedom. \textcolor{black}{More significantly, our findings provide valuable insight into the conditions under which magnon BEC can emerge in real spin systems with varying \textcolor{black}{interplanar} strengths and types of magnetic anisotropy.}

    \section*{Acknowledgments}
   The work at SKKU was supported by the National Research Foundation
    (NRF) of Korea (Grant no. RS-2023-00209121, 2020R1A5A1016518). S.M.K. and C.-L.H. are supported by the
    National Science and Technology Council in Taiwan with a
    Grant No. NSTC 114-2112-M-006-012. R. S. acknowledges the financial support provided by the Ministry of Science and Technology in Taiwan under Project No. NSTC-113-2124-M-001-003 and No. NSTC-113-2112M001-045-MY3,  as well as support from Academia Sinica for the budget of AS-iMATE11312. financial support from the Center of Atomic Initiative for New Materials (AIMat), National Taiwan University,  under Project No. 113L900801. 
    
   \section{DATA AVAILABILITY}
   The data that support the findings of the current study
   are available from the corresponding author upon reason
   able request.
\bibliography{KMMO}
\end{document}